# Proposition and validation of an original MAC layer with simultaneous accesses for low latency wireless control/command applications

Adrien van den Bossche, Thierry Val, Eric Campo

*University of Toulouse, UTM, LATTIS EA 4155 – SCSF group, Blagnac, France*
*{vandenbo, val, campo}@iut-blagnac.fr*

**Abstract:** Control/command processes require a transmission system with some characteristics like high reliability, low latency and strong guarantees on messages delivery. Concerning wire networks, field buses technologies like FIP offer this kind of service (periodic tasks, real time constraints…). Unfortunately, few wireless technologies can propose a communication system which respects such constraints. Indeed, wireless transmissions must deal with medium characteristics which make impossible the direct translation of mechanisms used with wire networks. The purpose of this paper is to present an original Medium Access Control (MAC) layer for a real time Low Power-Wireless Personal Area Network (LP-WPAN). The proposed MAC-layer has been validated by several complementary methods; in this paper, we focus on the specific Simultaneous Guaranteed Time Slot (SGTS) part.

## 1. INTRODUCTION

Today, wireless network technologies are widely used in many applications. Wireless eliminates expensive, heavy, not aesthetic cables, which are not easy to install or to use. Using a wireless network is sometimes a luxury, but it may be necessary in many cases of moving devices, like car tire sensors, embedded sensors on robots, etc. All these new needs encourage research and industrial to develop technologies and products in this domain.

A typical wireless sensor network (Culler *et al.* 2004) technology has to propose strong and reliable mechanisms for each level of the OSI model: Physical layer (PHY) must deal with poor Bit Error Rate, Medium Access Control layer (MAC) must avoid collisions and solve hidden terminal, Network layer (NWK) must enable automatic routing and insure reliability for mobile nodes (Badis *et al.,* 2004), and so on. For a wireless control/command application, a high reliability is required: the technology must propose some guarantees depending on the application (temporal bounding on transmission latency and packet forwarding, minimal throughput for critical nodes, maximal packet loses…). Adding Quality of Service (QoS) functionalities to the network is crucial in this type of real-time network application (Simplot-Ryl, 2005).

Our research works take place at the second level of the OSI-model (*link-layer*) for time-constrained and communicating applications such as robotics (van den Bossche *et al.* 2006) (van den Bossche *et al.* 2007) in not disturbed environments. The MAC sub-layer is in charge of the medium access organization, i.e. avoiding simultaneous transmissions which imply frame collisions and retransmissions, involving transmission latency. In the context of time-constrained wireless networks, we have proposed a new MAC method for the IEEE 802.15.4 wireless technology (van den Bossche *et al.* 2006) (van den Bossche *et al.* 2007); the proposed MAC enables guaranteed and periodic medium accesses thanks to a centralized scheduling and an exhaustive timeslot repartition between nodes. In this paper, we propose an improvement of the MAC in order to prevent the rarefaction of slots, by allowing nodes to access medium simultaneously (concept of SGTS – Simultaneous Guaranteed Time Slot, defined later).

After this introduction, we first present the IEEE 802.15.4 wireless network technology and the weaknesses we identified in the context of time-constrained networks. Then we propose a brief description of the new MAC and the SGTS improvement. At last, we present the validation by hardware prototyping and the obtained results of the SGTS concept are presented.

## 2. PRESENTATION OF IEEE 802.15.4

IEEE 802.15.4 standard (IEEE 2003) (IEEE 2006) proposes a two-layer protocol stack (physical-layer and data link-layer) for low power transceivers and low baud rate communications between embedded devices. Innovative concepts optimize energy saving. Moreover, IEEE 802.15.4 standard is promoted by the ZigBee Alliance (ZigBee Alliance, 2005) as the physical-layer and data-link-layer of the ZigBee Network specifications.

*2.1 Overview*

IEEE 802.15.4 proposes two PHY layers: PHY868/915 and PHY2450. The first one operates on both 868 MHz and 915 MHz radio bands. It proposes a very low data-rate (20 kbps at 868 MHz and 40 kbps at 915 MHz) with a simple BPSK modulation. The PHY2450 layer is more interesting: it allows a greater throughput (250 kbps) thanks to an O-QPSK modulation. Moreover thanks to its Direct Sequence Spread Spectrum (DSSS) coding, PHY2450 has excellent noise immunity (IEEE, 2003). The two PHY layers were designed for maximum energy saving: protocols are optimized for short and periodical data transfers. Nodes mostly stay in a "sleeping" mode called doze mode. Radio modem allows ultra low power consumption (40 µA) (Freescale Semiconductors, 2005) and nodes become operational in a very short time (330 µs). In doze mode, all radio functionalities are switched off, removing the ability to receive messages. The waking time has to be set before going in doze mode (synchronous wake-up) but sleeping devices may also wake-up if a local event occurs (asynchronous wake-up): motion detection for example.

*2.2 Medium Access Control (MAC) and topologies*

The standard IEEE 802.15.4 proposes two data link-layer topologies: Peer-to-Peer and Star. Peer-to-peer topology makes possible direct data transfers between devices in radio range on the same radio channel. Access to the medium uses the CSMA/CA protocol without RTS/CTS mechanism. On the contrary, Star topology needs a star coordinator: all data transfers go through the coordinator and messages are buffered during the dozing period. This functionality is called indirect data transfer. Star topology allows high energy saving thanks to an optimal distribution of sleeping periods between embedded devices. For synchronization, the star coordinator sends beacon frames. Inter-beacon period is called superframe. During the superframe, the nodes sleep until the next beacon, wake up and receive the beacon, ask the star coordinator for pending data, transmit and receive and then go to doze-mode again.

In addition to the classical CSMA/CA-based medium access method, IEEE 802.15.4 proposes a Contention Free method for the Star beaconed topology. Nodes can request for Guaranteed Time Slots (GTS) to the star coordinator. A GTS consists in one or several time slots dedicated to a particular node and cannot be used by other nodes. GTSs are announced by the beacon frame, a superframe contains up to seven GTS. The number of GTS reservations for a terminal node is directly linked to its communication bandwidth. This process of medium access reservation provides Quality of Service properties like bandwidth reservation or latency guarantees (Huang *et al.* 2006), like 802.11e HCF (IEEE, 2005).

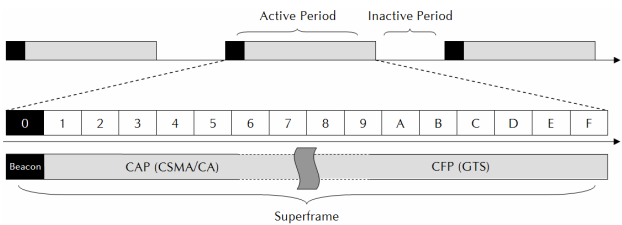

*Fig. 1: superframe structure*

The IEEE 802.15.4 superframe mixed structure (Fig. 1) combines both methods as follows: First, a star coordinator sends a beacon frame to indicate the network and coordinator addresses, the nodal data pending, the sizes of the Contention Access Period (CAP) and Contention Free Period (CFP). Then starts the CAP where the nodes and coordinator send/receive frames using CSMA/CA protocol. This time is also used for request from a node to obtain GTS in the next superframe. At the end of the CAP, the CFP starts as defined by the coordinator and broadcasted by the beacon. Medium access is possible only if the node has successfully obtained a GTS. At the end of the CFP, all nodes go to doze mode if not already and wait until the next beacon scheduled broadcast by using an internal wake up timer. This sleeping period is optional but greatly advised for energy saving.

$$BI = 15.36\ ms * 2^{BO} \text{ with } 0 \leq BO \leq 1 \qquad (1)$$

$$SFAP = 15.36\ ms * 2^{SO} \text{ with } 0 \leq SO \leq BO \qquad (2)$$

Therefore, the superframe is characterized by two temporal parameters *Beacon Order* (*BO*), *Superframe Order* (*SO*) announced in beacon frames: BO defines the time interval between two beacon messages. Beacon Interval (BI) is calculated as mentioned in (1). SO defines the SuperFrame "Active Portion" (SFAP = TCAP+TCFP) and is calculated as mentioned in (2). If BO and SO values are small, the network is more reactive (low latency) with lower energy saving. The greater is the difference between BO and SO, the more energy is saved. Thus, an appropriate value for these two parameters will be required considering the applications requirements.

## 3. IDENTIFIED WEAKNESSES OF THE MAC LAYER IN A LOW LATENCY APPLICATION CONTEXT

The medium access method proposed by the IEEE 802.15.4 standard is simple and flexible enough for auto-configured and spontaneous networks. The CSMA/CA protocol enables automatic adaptation of the medium access, even if the number of nodes is important. However, if an IEEE 802.15.4 network is used in a time-constrained or real-time context, the CSMA/CA based MAC does not fit because it does not propose any guarantee on messages delivery, since it is a *best effort* protocol. As shown in the above section, 802.15.4 adopts an interesting mechanism of medium access reservation (GTS) to free some privileged nodes from the collision phenomenon. However, the medium reservation is conditioned by two factors: First, the network must be maintained within its capacity and avoid saturation (Misic *et al.* 2006). Unfortunately, the standard does not grant to a star coordinator the capability to permanently maintain some bandwidth for a particular node. The GTS reservation process works as "first come, first served" and is not an acceptable rule of distribution in terms of Quality of Service. Second, the primitive call "GTS.request" generates a message sent to the star coordinator during the CAP using the CSMA/CA protocol. As this protocol is Best-effort, it can not provide any temporal guarantee. Many contributions have been proposed in order to insure QoS functionalities at MAC-level by an optimization of beacon or GTS scheduling, for example (Koubâa *et al.* 2007) or (Francomme *et al.* 2007). In this paper, the work is focussed on the proposition of a full deterministic MAC by using a new scheduling.

To achieve a communication between devices in an application with temporal constraints, it is essential to insure bandwidth and network latency for a number of known critical nodes. Moreover, the different devices may have different communication requirements: strong sporadic flows, regular flows with time dependency, etc. In this time-critical context, the standard IEEE 802.15.4 has others weaknesses:

- The GTS frequency is based upon the superframe frequency, the star coordinator *BO* and *SO* internal clock parameters. The nodes may only need to communicate from time to time and not on a regular base. In other words, it is extremely difficult for sensors with different data communication need to cohabit on the same star without loss of continuity and optimal GTS distribution.
- If several stars are in the same radio range and on the same channel, there is a high probability for collisions even during the CFP because the standard does not provide communication protocols between star coordinators.

According to all these observations, the mechanism of medium reservation could be greatly improved by:

- A fully deterministic access method to insure GTS for selected known nodes at each superframe,
- A more flexible GTS allocation to support various access frequencies and bandwidth,
- The introduction of a new protocol between star coordinators in order to avoid GTS collisions.

## 4. NEW FUNCTIONALITIES PROPOSED BY THE MAC

In (van den Bossche *et al.* 2006) (van den Bossche *et al.* 2007), we have presented an original MAC layer for a small time-constrained sensor network, based on the IEEE 802.15.4 Guaranteed Time Slot mechanism. The proposed medium access method enables periodical tasks to access medium at predetermined and regular moments. Moreover, the new GTSs are guaranteed *beyond the star*, in order to avoid GTS collisions between stars in the same radio range and channel. The principal functionalities of the new MAC are presented in the next paragraphs.

*4.1 New proposed functionalities*

The proposed mechanisms are intended to achieve the following new functionalities:

- With the present IEEE 802.15.4 standard, only nodes can request for a GTS. We propose to give a star coordinator the ability to allocate GTS at any time to any known node, in anticipation of the request, depending on the application needs. This functionality makes possible deterministic network associations for critical nodes (i.e. in a bounded time). Indeed, a mobile node might be capable of changing its coordinator without loosing its guaranteed medium access. We call this ability PDS, for *Previously Dedicated Slot*.
- With the present standard, the GTSs were managed by the coordinator with internal messages within the star. We now propose a mechanism to extend these communications between coordinators to avoid "GTS collisions" (two coordinators give a same GTS for two nodes by two different stars in the same radio range).
- With the present standard, the GTSs were placed in the CFP, at the end of the superframe. We propose that the GTSs will be laid out anywhere in the superframe at the coordinator discretion. This functionality will enable us to optimize GTS distribution and with a possible extension to generate an optimized global superframe composed of several stars, even if CSMA performance will be sacrificed because CAP periods are non-contiguous.
- With the present standard, a GTS appeared in every single superframe after allocation. We now suggest regulating this GTS inclusion in the superframe at a lower frequency to fit the node needs. Thus, a GTS can appear in one superframe out of two, one out of four, one out of eight, etc. We introduce the notion of reservation level *n*, an integer from 0 to $n_{MAX}$. The GTS of a node with a reservation level *n* will appear in every $p = 2^n$ superframes. It will allow different QoS traffics to cohabit within the same star without need for adjustments of *BO* and *SO* parameters.

*4.2 Aimed network topology and role of the PAN-coordinator*

The proposed MAC organization is managed by the PAN coordinator. This central entity:

- broadcasts *superbeacon* frames to synchronize all the network devices. This broadcasting is mandatory to provide a global synchronization over the network and reduce frame collision (Rowe *et al.*, 2006),
- receives all GTS request messages,
- schedules the medium accesses of critical nodes, i.e. nodes which request GTS.

Thanks to this centralized organisation, the collision GTS phenomenon is limited. Now, a GTS is a "real Guaranteed Time Slot", since no other node is allowed to transmit during the timeslot, even if the GTS is dedicated to another star node. Of course, *best-effort* medium access using CSMA/CA is still possible for non critical nodes. The aimed network topology is illustrated on figure 2.

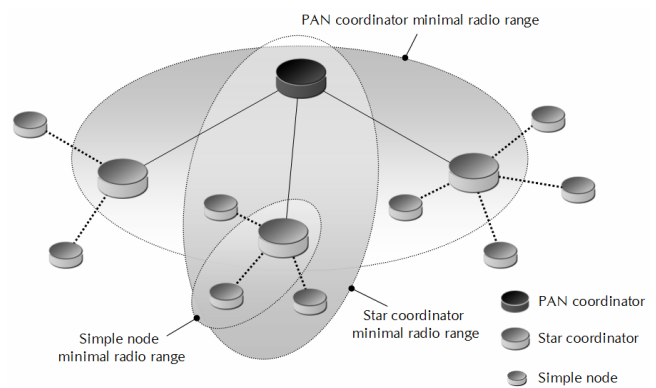

*Fig. 2: Network topology and minimal radio ranges between nodes*

The proposed network is composed of three node types: a unique PAN-coordinator, one or several star coordinators (one for each star) and one or several simple nodes. Each simple node must be associated to a star coordinator. In a first approach, we have considered that each star coordinator is in the radio range of the PAN-coordinator; this hypothesis is not so constraining thanks to the availability of high-power IEEE 802.15.4 devices such as MaxStream XBeePRO (MaxStream 2006) which enable extended radio ranges up to 1 mile. In return, a star coordinator may not be in the range of all others star coordinators – the PAN coordinator solves the *hidden star coordinator issue* by distributing slots for coordinator beacons, GBS (*Guaranteed Beacon Slots*).

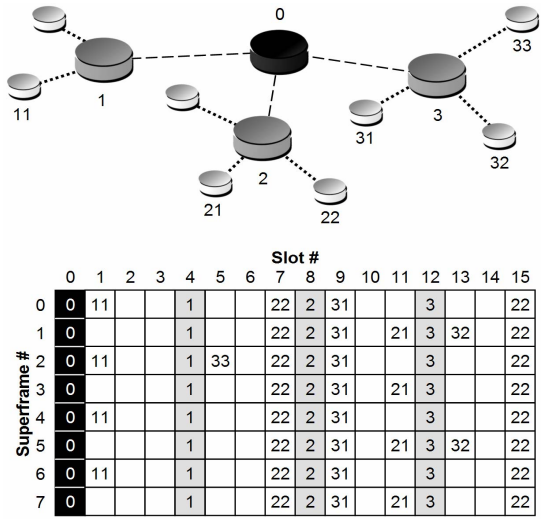

Fig.3: An example of scheduling by PAN-coordinator

The figure 3 illustrates a medium scheduling with $n_{MAX} = 3$. The table in the figure 3 shows the timeslot repartition decided by the PAN-coordinator for the next $2^{n_{MAX}}$ future superframes. The PAN-coordinator broadcasts its *superbeacons* on each slot #0. Slots #4, #8 and #12 are GBS for the three coordinators (nodes 1, 2, 3). Simple nodes (11, 21, 22, 31, 32 and 33) have different needs, so they have requested one or more GTS with different reservation level $n$ (one superframe out of eight for node 33, one superframe out of four for node 32, one superframe out of two for nodes 11 and 21 and at last a GTS in each superframe for nodes 22 and 31).

## 5. PRESENTATION OF THE SGTS CONCEPT

In section 4.2, we illustrated the scheduling organized by the PAN-coordinator. To prevent the rarefaction of slots, we propose a possibility, for the PAN-coordinator, to give the same timeslot for two different star nodes if the radio conditions make it possible, enabling the possibility for each node to transmit its message at the same time without any collision. This optional functionality of spatial reusing, as in (Lee *et al.* 2006), should increase the performances of the network since two distant nodes can send their message at the same time. Nevertheless, this functionality may be used carefully: the GTSs must keep their "Guaranteed" characteristic. In a first approach, we consider that a SGTS will regroup in a single timeslot the transmission of no more than two transmitters.

In order to determine if two nodes can transmit at the same time without making a perturbation on the transmission (i.e. if two allocated GTS can be regrouped in a single SGTS), the PAN-coordinator must know the radio reception conditions of the destination nodes. Indeed, the SGTS needs to be negotiated by taking into account the receiver opinions. The PAN-coordinator must take into account only the two destination nodes to decide the SGTS attribution. If we consider two node pairs (two transmitters, two receivers), the two receivers must not be perturbed by the other transmitter, as shown on figure 4. On this illustration, A2 sends a message to A1 and B2 sends a message to B1. The SGTS can be negotiated, i.e. A2 and B2 can both transmit their message at the same time, only if A1 is not perturbed by B2 and B1 is not perturbed by A2. A *non perturbation threshold,* in dB, based on the two signal levels (RSSI, *Received Signal Strength Indicator*) in the two receivers, must be identified. This threshold has been evaluated by a study on real prototype and results are presented in the next section.

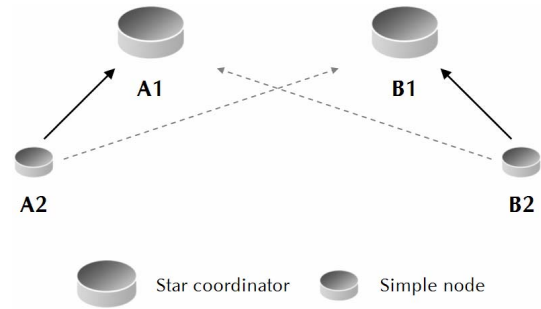

Fig. 4: Illustration of the non-perturbation principle to negotiate a SGTS between two node pairs

Nevertheless, each SGTS must be negotiated with precaution, particularly if the concerned nodes are mobile nodes. In this case, we recommend disabling this functionality.

## 6. SGTS VALIDATION BY HARDWARE PROTOTYPING

The concept of Simultaneous GTS needs to be validated and the *non perturbation threshold* has to be identified. The real prototyping seems to be the best way to validate the SGTS concept. Therefore we have developed a prototype of the MAC layer and deployed a network based on a couple of FREESCALE™ IEEE 802.15.4 devices (Freescale Semiconductors, 2005); this type of 802.15.4 devices is totally reprogrammable, which allowed us to implement the proposed MAC layer and the SGTS negotiation. This prototype also enables us to evaluate the non perturbation threshold evoked in the section 5.

### 6.1 The prototype network and its topology

The prototype network is composed of five nodes: a PAN-coordinator (PC), two star coordinators (C1, C2) and two simple nodes (N1, N2). Each simple node is associated to a different coordinator. The network topology is shown on figure 5 while the superframe structure is represented on figure 6.

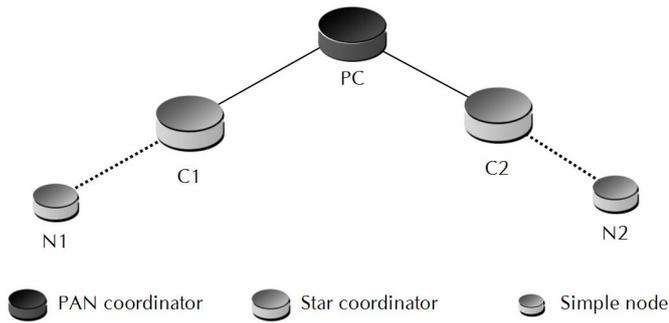

Fig.5: The prototype network topology

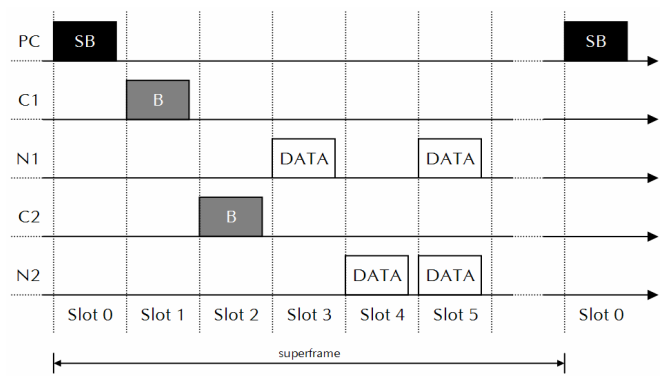

Fig. 6: Superframe structure during the tests

For this study case, we consider that both N1 and N2 have obtained a GTS and can freely use it to send data messages to their star coordinator (N1 to C1 in slot #3 and N2 to C2 in slot #4). Slot #5 is used by both N1 and N2. In this study, the objective is to evaluate the *non perturbation threshold*, i.e. to measure the number of collisions in slot #5. In order to evaluate the perturbation of the other transmitter (N2 for C1 and N1 for C2), each coordinator listens the messages sent by the two nodes and gets the RSSI value during the slots #3 and #4; the RSSI difference is calculated at the end of slot #4 if both messages from N1 and N2 where received. In slot #5, each coordinator listens to the medium; if the coordinator receives the message sent by its node, the *result is positive*. If the coordinator receives the message of the other node *or* a collision, the *result is negative*. Note that on the FREESCALE[TM] IEEE 802.15.4 devices used, the transmit power can be adjusted from -16dBm up to +3.6dBm; this functionality enables us to implement an automatic variation of node transmitting power to increase the measure range without moving the nodes. All measures have been realized into an anechoic chamber, i.e. without any noise.

*6.2 Obtained results*

The results obtained on figure 7 are really interesting: in most cases, two transmissions can be done at the same time without perturbing the other receiver. In fact, measures show (a well-known result) that there is only a 10 dB window where the SGTS should not be negotiated because of an important risk of collision.

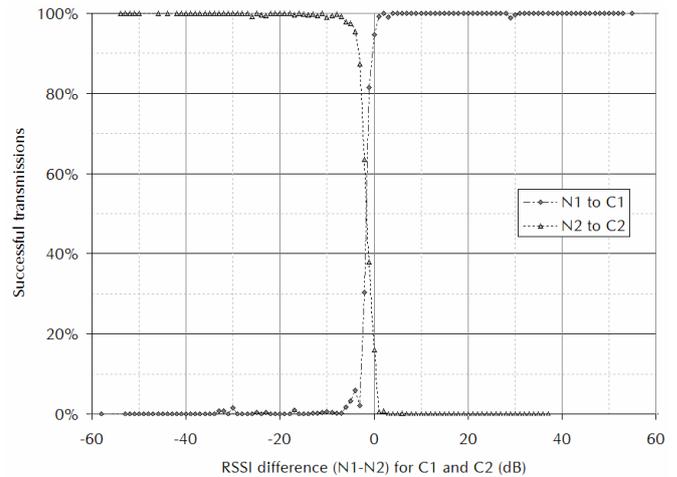

Fig.7: SGTS validation by hardware prototyping: obtained results

Moreover, the results presented on figure 7 show another interesting point: the two line plots do not cross at 0 dB. It indicates a certain inequality between the two nodes: we notice that the node N1 has a greater probability of being received by the two receivers, even if its message is received with a smaller RSSI than the message of N2. The same results have been obtained with other measures. Indeed, a study on synchronisation has shown us that our three-level synchronisation procedure (PAN-coordinator, star coordinator, simple node) of our prototype network was not so perfect: one of the two nodes takes advantage on the other by having a little temporal advance (few μs). We have made a cross-comparison of these two studies and determined that the favourite node in the SGTS study is the one which takes the temporal advance in the synchronisation study.

In order to verify this hypothesis, we make others measures with the same synchronisation for the two nodes, as illustrated on figure 8. Of course, this study is unusable for transmitting data, since the destination of the two message sent at the same time is the same coordinator.

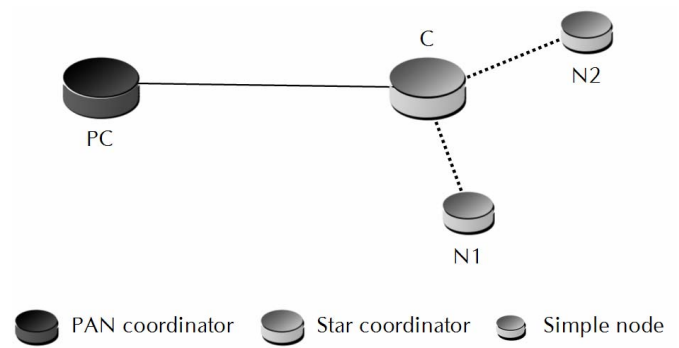

Fig. 8: Network topology of the second study

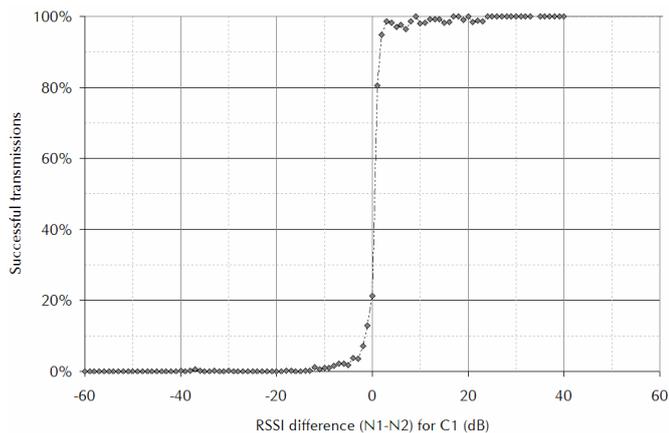

*Fig.9: Obtained results with the second SGTS study*

The obtained results, on figure 9, show that the node equality is now correct. The line plots cross now at 0 dB.

## 6. CONCLUSIONS

Our works deal with a new MAC-layer for a LP-WPAN IEEE 802.15.4 with real deterministic capabilities. Thanks to this medium access method, time-constrained nodes can negotiate a periodical and guaranteed medium access, as required in a control/command application. Moreover, the reservation level parameter ($n$) enables the cohabitation of different profiles of traffics in the same network without changing *BO* and *SO*. In this paper, we have presented an improvement of this MAC-layer which enables simultaneous accesses by the use of specifically described SGTS. SGTSs have been implemented and tested on a couple of hardware devices. Measures have proved the possibility of simultaneous transmissions without collision when propagation conditions are accepted by the PAN-coordinator. On theses first results, we are very optimistic about the potential improvement of the performances of new MAC-layer for a time-constrained wireless network. The future works concern the optimisation of the scheduling, including the optional functionality of SGTS. We are participating in a national project (OCARI 2007) focussed on the development of a low power and large scale sensor network with time constrained messages delivery for industrial applications.

## REFERENCES

Badis, H. and Al Agha, K (2004). QOLSR Multi-path Routing for Mobile Ad Hoc Networks Based on Multiple Metrics. *59th IEEE Vehicular Technology Conference (IEEE VTC'04-Spring)*, Milan, Italy

Culler, D., Estrin, D. and Srivastava, M. (2004). Overview of Sensor Networks. *IEEE Computer* **vol. 37, issue 8.**

Francomme J., Mercier G. and Val T. (2007). Beacon Synchronization for GTS Collision Avoidance in an IEEE 802.15.4 Meshed Network, IFAC FET'07, Toulouse, France.

Freescale Semiconductors (2005). MC13192 2.4 GHz Low Power Transceiver for the IEEE 802.15.4 Standard, Reference Manual Document, MC13192RM Rev. 1.3.

Huang, Y.K., Pang, A.C., Kuo, T.W. (2006). AGA: Adaptive GTS Allocation with Low Latency and Fairness Considerations for IEEE 802.15.4. *IEEE International Conference on Communications* (*ICC 2006*), Istanbul, Turkey.

Koubâa A., Cunhà A. and Alves M. (2007), A Time Division Beacon Scheduling Mechanism for IEEE 802.15.4 / Zigbee Cluster-Tree Wireless Sensor Networks, ECRTS'2007, Pisa, Italy.

LAN-MAN Standards Committee of the IEEE Computer Society (2003). Part 15.4: Wireless Medium Access Control (MAC) and Physical Layer (PHY) Specifications for Low-Rate Wireless Personal Area Networks (LR-WPANs)

LAN-MAN Standards Committee of the IEEE Computer Society (2005). Part 11: Wireless LAN Medium Access Control (MAC) and Physical Layer (PHY) Amendment: Medium Access Control (MAC) Quality of Service (QoS) Enhancements

LAN-MAN Standards Committee of the IEEE Computer Society (2006). Part 15.4b: Wireless Medium Access Control (MAC) and Physical Layer (PHY) Specifications for Low-Rate Wireless Personal Area Networks (LR-WPANs)

Lee W. L., Datta A. and R.Cardell-Oliver (2006), FlexiMAC: A flexible TDMA-based MAC protocol for fault-tolerant and energy-efficient wireless sensor networks. *14th IEEE International Conference on Networks (ICON'06)*, **vol. 2, pp 1-6**, Singapore.

MaxStream, Inc, (2006). IEEE 802.15.4/ZigBee XBEEPRO RF Modules datasheet.

Misic, J., Shafi, S. and Misic, V (2006). Real-time admission control in 802.15.4 sensor clusters: *International Journal of Sensor Networks* **vol. 1, issue 1/2** pp 34-40

OCARI (2007). The OCARI project web site http://ocari.lri.fr

Rowe A., Mangharam R. and Rajkumar R (2006). RT-Link: A Time-Synchronized Link Protocol for Energy Constrained Multi-hop Wireless Networks. *IEEE International Conference on Sensors, Mesh and Ad Hoc Communications and Networks (IEEE SECON)*, Reston, VA.

Simplot-Ryl, D. (2005). Real-time aspects in Wireless Sensor Networks. *ETR'05*, Nancy, France

van den Bossche, A., Val, T. and Campo, E. (2006). Proposition of a full deterministic medium access method for wireless network in a robotic application. *IEEE VTC'06-Spring: Vehicular Technology Conference*, Melbourne, Australia

van den Bossche, A., Val, T. and Campo, E (2007). Prototyping and performance analysis of a QoS MAC layer for industrial wireless network, IFAC FET'07, Toulouse, France.

ZigBee Alliance (2005). ZigBee Specification